\documentclass[12pt]{article}
\usepackage{amssymb}
\usepackage{amsmath}
\usepackage{epsf}
\usepackage{epsfig}
\usepackage{cite}
\usepackage[latin1]{inputenc}
\newcommand{\lsim}{
\mathrel{\hbox{\rlap{\hbox{\lower4pt\hbox{$\sim$}}}\hbox{$<$}}}}

\newcommand{\gsim}{
\mathrel{\hbox{\rlap{\hbox{\lower4pt\hbox{$\sim$}}}\hbox{$>$}}}}

\def\Adel{\mathcal{A}_{\Delta\Gamma}}

\def\D0{D\O }
\def\mixAng{\varphi_M}

\begin{document}
\begin{titlepage}
\vspace*{-1.3truecm}

\begin{flushright}
Nikhef-2011-028 \\
DSF/11/2011
\end{flushright}

\vspace*{1.8truecm}

\begin{center}
\boldmath
{\Large{\bf Exploring CP Violation and $\eta$--$\eta'$ Mixing\\ 
with the $B^0_{s,d} \to J/\psi \eta^{(\prime)}$ Systems}}
\unboldmath
\end{center}

\vspace{0.9truecm}

\begin{center}
{\bf Robert Fleischer,\,${}^a$ Robert Knegjens\,${}^a$ \,and\, Giulia Ricciardi\,${}^{b}$}

\vspace{0.5truecm}

${}^a${\sl Nikhef, Science Park 105,
NL-1098 XG Amsterdam, The Netherlands}

\vspace{0.3truecm}

${}^b${\sl Dipartimento di Scienze Fisiche, Universit\`a di Napoli Federico II  and  
I.N.F.N., Sezione di Napoli, Complesso Universitario di Monte Sant'Angelo, 
Via Cintia, I-80126~Napoli, Italy}

\end{center}

\vspace{1.6cm}
\begin{abstract}
\vspace{0.2cm}\noindent
The $B^0_{s,d} \to J/\psi \eta^{(\prime)}$  decays provide new terrain 
for exploring CP violation. After briefly discussing $\eta$--$\eta'$ mixing, we analyse the 
effective lifetimes and CP-violating observables of the $B_s$ channels, which allow us to 
probe New-Physics effects in $B^0_s$--$\bar B^0_s$ mixing. We have a critical look at these 
observables and show how hadronic corrections can be controlled by means of the $B_d$ 
decays. Using measurements of the $B^0_{s,d}\to J/\psi \eta^{(\prime)}$ branching ratios by 
the Belle collaboration, we discuss tests of the $SU(3)_{\rm F}$ flavour symmetry of strong 
interactions, obtain the first constraints on the hadronic parameters entering the 
$B^0_{s,d} \to J/\psi \eta$ system, and predict the $B^0_d\to J/\psi \eta'$ branching ratio at the 
$5\times10^{-6}$ level. Furthermore, we present strategies for the determination of the 
$\eta$--$\eta'$ mixing parameters from the $B^0_{s,d} \to J/\psi \eta^{(\prime)}$ observables. 
We also observe that the $B^0_{s,d} \to J/\psi \eta$ and $B^0_{s,d} \to J/\psi \eta'$ decays 
are -- from a formal point of view -- analogous to the quark--antiquark and tetraquark descriptions 
of the $f_0(980)$ in the $B^0_{s,d} \to J/\psi f_0(980)$ channels, respectively. 
\end{abstract}

\vspace*{0.5truecm}
\vfill
\noindent
October 2011
\vspace*{0.5truecm}

\end{titlepage}

\thispagestyle{empty}
\vbox{}
\newpage

\setcounter{page}{1}

\section{Introduction}
With the Large Hadron Collider (LHC) at CERN collecting plenty of data, tests of the 
Standard Model (SM) have entered a new era. Concerning the exploration of the 
quark-flavour sector, the study of CP violation in $B_s$-meson decays at the LHCb 
experiment is one of the most exciting aspects of this endeavor.  In addition to 
various, by now ``standard", $B_s$ decays  \cite{bib:LHCbRoadMap},  another 
interesting probe is offered by the $B^0_s\to J/\psi \eta^{(\prime)}$ channels. 
In these decays, New Physics (NP) can enter through CP-violating contributions to 
$B^0_s$--$\bar B^0_s$ mixing, which is a strongly suppressed loop phenomenon in the 
SM (see, for instance, Refs.~\cite{Bmix} and references therein). 

In Ref.~\cite{skands}, a determination of the angle $\gamma$ of the 
Unitarity Triangle (UT) of the Cabibbo--Kobayashi--Maskawa (CKM) matrix was 
proposed that relates the $B^0_d\to J/\psi \eta^{(\prime)}$ decays to 
$B^0_s\to J/\psi \eta^{(\prime)}$ through the $SU(3)_{\rm F}$ flavour symmetry
of strong interactions. This method is a variant of the $B^0_{s,d}\to J/\psi K_{\rm S}$
strategy proposed in Ref.~\cite{RF-BpsiK}.  As was recently shown \cite{DeBFK}, the 
extraction of $\gamma$ from the $B^0_{s,d}\to J/\psi K_{\rm S}$ system is possible
at LHCb but cannot compete with other strategies \cite{bib:LHCbRoadMap}; the 
situation for $B^0_{d,s}\to J/\psi \eta^{(\prime)}$ looks even more challenging. 
However, the $B^0_s\to J/\psi K_{\rm S}$ mode will still play an important role at
LHCb as a ``control channel", allowing us to take hadronic SM corrections in the 
extraction of the UT angle $\beta$ from the CP violation in $B^0_d\to J/\psi K_{\rm S}$ 
into account \cite{RF-BpsiK,DeBFK}.

In the present paper, we shall follow a similar avenue, assuming that $\gamma$ will be 
measured at LHCb with a precision of a few degrees by means of  the corresponding 
benchmark decays in the next couple of years \cite{bib:LHCbRoadMap}. We will then 
use the $B^0_d\to J/\psi \eta^{(\prime)}$ decays to control the hadronic corrections to 
observables of the $B^0_s\to J/\psi \eta^{(\prime)}$ modes, which are sensitive to
NP effects in $B^0_s$--$\bar B^0_s$ mixing. To be specific, we will study the 
effective lifetimes and CP-violating rate asymmetries of the $B^0_s\to J/\psi \eta^{(\prime)}$ 
decays. In contrast to the rate asymmetries, the lifetime analysis utilizes the sizable 
width difference $\Delta\Gamma_s$ between the $B_s$ mass eigenstates and 
requires only untagged $B_s$ data samples. Here one does not distinguish 
between initially present $B^0_s$ or $\bar B^0_s$ mesons, which is advantageous
from an experimental point of view.

The power of the $B^0_s\to J/\psi \eta^{(\prime)}$ observables to reveal CP-violating 
NP contributions to $B^0_s$--$\bar B^0_s$ mixing is limited by doubly Cabibbo-suppressed 
contributions to the decay amplitudes. We will explore the impact of the relevant 
non-perturbative parameters, which cannot be calculated reliably in QCD, 
and discuss how they can be determined from $B^0_d\to J/\psi \eta^{(\prime)}$ data 
through  the $SU(3)_{\rm F}$ flavour symmetry of strong interactions. Thanks to their 
different CKM amplitude structure, the hadronic parameters are not doubly 
Cabibbo-suppressed in these channels, thereby leading to significant effects in the 
corresponding observables.

Using $B^0_{s,d}\to J/\psi \eta$ branching ratio measurements by the Belle collaboration, 
we introduce quantities to probe $SU(3)_{\rm F}$-breaking effects, obtain first constraints 
on the relevant penguin parameters, and discuss strategies to extract the $\eta$--$\eta'$ 
mixing parameters. We also point out that the $B^0_{s,d}\to J/\psi \eta$ and 
$B^0_{s,d}\to J/\psi \eta'$ decays are -- from a formal point of view -- analogous to the 
quark--antiquark and tetraquark descriptions of the scalar $f_0(980)$ state in 
$B^0_s\to J/\psi f_0(980)$, respectively \cite{FKR}. For simplicity, we shall from here on 
abbreviate the $f_0(980)$ as $f_0$.

Unfortunately, the experimental analyses of the $B^0_s\to J/\psi \eta^{(\prime)}$ modes 
are complicated by the reconstruction of the $\eta^{(\prime)}$ decays. The most prominent 
channels are $\eta\to 2\gamma$, $3\pi^0$, $\pi^+\pi^-\pi^0$, $\pi^+\pi^-\gamma$ 
and $\eta'\to \pi^+\pi^-\eta$, $\rho^0\gamma$, $\pi^0\pi^0\eta$ \cite{PDG}, which
have challenging signatures for studies at hadron colliders. At the future $e^+e^-$ 
SuperKEKB and SuperB projects the prospects of measuring these decays may
be more promising.

The outline is as follows: in Section~\ref{sec:ampl}, we give a brief overview of 
$\eta$--$\eta'$ mixing and discuss how it is implemented in the $B^0_s\to J/\psi \eta^{(\prime)}$ 
decay amplitudes. In Section~\ref{sec:obs}, we turn to the effective lifetimes and 
the CP-violating observables of the $B^0_s\to J/\psi \eta^{(\prime)}$ transitions. 
In Section~\ref{sec:contr}, we focus on the $B^0_d\to J/\psi \eta^{(\prime)}$ decays and
their role as control channels. Finally, we discuss determinations of the $\eta$--$\eta'$
mixing parameters from the $B^0_{s,d}\to J/\psi \eta^{(\prime)}$ branching ratios in
Section~\ref{sec:mix}, and summarize our conclusions in Section~\ref{sec:concl}.

\boldmath
\section{The $B^0_s\to J/\psi \eta^{(\prime)}$ Decay Amplitudes}\label{sec:ampl}
\unboldmath
Before focusing on the $B^0_s\to J/\psi \eta^{(\prime)}$ decays, we will first give a brief overview 
of $\eta$--$\eta'$ mixing. The physical $|\eta\rangle$ and $|\eta'\rangle$ states are mixtures of the 
octet and singlet states $|\eta_8\rangle$ and $|\eta_1\rangle$, respectively, and can be written as 
follows \cite{PDG}:
\begin{equation}
\left(\begin{array}{c}
|\eta\rangle\\
|\eta'\rangle
\end{array}
\right)=
\left(\begin{array}{cc}
\cos\theta_P & -\sin\theta_P\\
\sin\theta_P & \cos\theta_P
\end{array}
\right)\cdot
\left(\begin{array}{c}
|\eta_8\rangle\\
|\eta_1\rangle
\end{array}
\right),
\end{equation}
where
\begin{equation}
|\eta_8\rangle=\frac{1}{\sqrt{6}}\left(|u\bar u\rangle + |d\bar d\rangle - 2 |s\bar s\rangle\right), \quad
|\eta_1\rangle=\frac{1}{\sqrt{3}}\left(|u\bar u\rangle + |d\bar d\rangle + |s\bar s\rangle \right).
\end{equation}
The mixing between the octet and singlet states is a manifestation of the breaking of the 
$SU(3)_{\rm F}$ flavour symmetry of strong interactions. Alternatively, $\eta$--$\eta'$ mixing 
can be described in terms of the isospin singlet states 
\begin{equation}\label{mixing1}
|\eta_q\rangle \equiv\frac{1}{\sqrt{2}}\left(|u\bar u\rangle + |d\bar d\rangle\right), \quad
|\eta_s\rangle\equiv|s\bar s\rangle.
\end{equation} 
By also taking the possible mixing with a purely gluonic component $|gg\rangle$ into 
account, we can write the following expressions (for a recent detailed discussion, 
see Ref.~\cite{DiDonato:2011kr}):
\begin{eqnarray}
|\eta\rangle&=&\cos\phi_P|\eta_q\rangle - \sin\phi_P|\eta_s\rangle\label{eta-1},\\
|\eta'\rangle &=& \cos\phi_G\sin\phi_P|\eta_q\rangle + \cos\phi_G\cos\phi_P|\eta_s\rangle
+\sin\phi_G|gg\rangle\label{eta-2}.
\end{eqnarray}
Here it has been assumed, for simplicity, that the heavier $\eta^\prime$ contains a larger 
gluonic admixture than the lighter $\eta$ and that the coupling of the latter state 
to $|gg\rangle$ is negligible. Estimates give $\sin^2 \phi_G \sim 0.1$~\cite{Ambrosino:2009sc}, 
i.e.\ $|\phi_G|\sim 20^\circ$, which indicates that the impact of this contribution is suppressed.

The mixing angle $\phi_P$ is still subject of ongoing studies, using data for processes
such as $D_s^+ \to \eta^{(\prime)}\ell^+\nu_\ell$ decays and the two-photon width of 
the $\eta^{(\prime)}$ mesons (see Ref.~\cite{DiDonato:2011kr} and references therein). 
The full spectrum of results correspond to $30^\circ\lsim \phi_P\lsim 45^\circ$, with 
the majority of analyses converging at values of $\phi_P$ around  $40^\circ$. Consequently, 
the relations
\begin{equation}\label{ang-rel}
\cos\phi_P \approx  \sqrt{\frac{2}{3}},
\quad
\sin\phi_P \approx \sqrt{\frac{1}{3}},
\end{equation} 
where the numerical values correspond to $\phi_P=35^\circ$, are affected by uncertainties 
of ${\cal O}(20\%)$. These approximate relations result in the simple expressions 
\begin{equation}\label{eta-wf}
|\eta\rangle\approx \frac{1}{\sqrt{3}}\left( |u\bar u\rangle + |d\bar d\rangle - |s\bar s\rangle\right)
\end{equation}
\begin{equation}\label{etap-simple}
|\eta'\rangle\approx \frac{1}{\sqrt{6}}\left( |u\bar u\rangle + |d\bar d\rangle + 
2 |s\bar s\rangle\right)\cos\phi_G + \sin\phi_G|gg\rangle,
\end{equation}
which are useful for $SU(3)_{\rm F}$ analyses of non-leptonic $B$-meson 
decays with $\eta^{(\prime)}$ mesons in the final states \cite{DGR,CGR}. In our study
we shall follow a similar conceptual avenue, keeping, however, $\phi_P$ as a free parameter.

The $B^0_s\to J/\psi \eta$ mode has dynamics very similar to $B^0_s\to J/\psi f_0$  
with a quark--antiquark description assumed for the $f_0$~\cite{FKR}. In particular, the decay 
topologies are the same, and to obtain the transition amplitude we only need to make the 
following substitutions in the relevant formulae:
\begin{equation}
\cos\varphi_{\rm M} \to -\sin\phi_P, \quad \sin\varphi_{\rm M} \to \cos\phi_P,
\end{equation}
i.e.\  $\varphi_{\rm M}$ introduced in Ref.~\cite{FKR} should be replaced by 
$\phi_P+90^\circ$.  The $f_0$ mixing angle corresponding to \eqref{ang-rel}, 
$\varphi_{\rm M}\approx125^\circ$, is consistent with phenomenological analyses 
of the scalar $f_0$ state in the quark--antiquark picture (see Ref.~\cite{FKR} and 
references therein).

Using the unitarity of the CKM matrix, the decay amplitude can be written as
\begin{equation}\label{ampl-1}
A(B^0_s\to J/\psi \eta)=\left(1-\frac{\lambda^2}{2}\right){\cal A}_\eta 
\left [1+\epsilon \, b_\eta  e^{i\vartheta_\eta } e^{i\gamma}  \right]
\end{equation}
with
\begin{equation}\label{A-expr}
{\cal A}_\eta = -\lambda^2 A \left[ \sin\phi_P\left\{\tilde{A}^{(c)}_{\rm T} + \tilde{A}^{(ct)}_{\rm P} + 
\tilde{A}^{(c)}_{\rm E} + \tilde
	{A}^{(ct)}_{\rm PA}\right\}
	-\frac{1}{\sqrt{2}}\cos\phi_P \left\{ 2 \tilde{A}^{(c)}_{\rm E} +
	2 \tilde{A}^{(ct)}_{\rm PA}\right\} \right]
\end{equation}
and
\begin{equation}\label{b-expr}
b_\eta  e^{i\vartheta_\eta } = R_b\left[
	\frac{\sin\phi_P\left\{\tilde{A}^{(ut)}_{\rm P} + \tilde{A}^{(ut)}_{\rm PA}\right\} -
	\frac{1}{\sqrt{2}}\cos\phi_P \left\{\tilde{A}^{(u)}_{\rm E} + 2 \tilde{A}^{(ut)}_{\rm PA}\right\} }
	{\sin\phi_P\left\{\tilde{A}^{(c)}_{\rm T} + \tilde{A}^{(ct)}_{\rm P} + \tilde{A}^{(c)}_{\rm E} + \tilde
	{A}^{(ct)}_{\rm PA}\right\}
	-\frac{1}{\sqrt{2}}\cos\phi_P \left\{ 2 \tilde{A}^{(c)}_{\rm E} +2 \tilde{A}^{(ct)}_{\rm PA}\right\}}
	\right],
\end{equation}
where we have used the isospin and $SU(3)_{\rm F}$ flavour symmetries of strong interactions 
to identify certain amplitudes and hereby simplify the expressions. In analogy to the
discussion in Ref.~\cite{FKR}, ${\cal A}_\eta$ and $b_\eta  e^{i\vartheta_\eta }$ are 
CP-conserving strong parameters, which encode the hadron dynamics of the 
$B^0_s\to J/\psi \eta$ decay; the labels T, P, E and PA refer to tree, penguin, exchange 
and penguin annihilation topologies, respectively. As usual, 
$\lambda\equiv|V_{us}|=0.2252 \pm 0.0009$ denotes the Wolfenstein parameter \cite{PDG}, while 
\begin{equation} \label{VubVcb}
\epsilon\equiv\frac{\lambda^2}{1-\lambda^2}=0.0534  \pm 0.0005, \quad
A\equiv \frac{|V_{cb}|}{\lambda^2}\sim0.8, \quad
R_b\equiv \left(1-\frac{\lambda^2}{2}\right)\frac{1}{\lambda}\left|\frac{V_{ub}}{V_{cb}}\right|\sim0.5.
\end{equation}
Using (\ref{ang-rel}), we obtain the following simplified expressions:
\begin{equation}\label{A-eta}
{\cal A}_\eta \approx  -\lambda^2 A \sqrt{\frac{1}{3}}\left[\tilde A^{(c)}_{\rm T} + \tilde A^{(ct)}_{\rm P} - 
\tilde A^{(c)}_{\rm E} - \tilde A^{(ct)}_{\rm PA}\right],
\end{equation}
\begin{equation}\label{b-eta}
b_\eta e^{i\vartheta_\eta} \approx R_b\left[
	\frac{\tilde A^{(ut)}_{\rm P} -\tilde A^{(u)}_{\rm E} - \tilde A^{(ut)}_{\rm PA}}
	{\tilde A^{(c)}_{\rm T} + \tilde A^{(ct)}_{\rm P} - \tilde A^{(c)}_{\rm E} - 
	\tilde A^{(ct)}_{\rm PA}}\right].
\end{equation}

The $B^0_s\to J/\psi \eta'$ amplitude takes the same form as
(\ref{ampl-1}). The corresponding parameters ${\cal A}_{\eta'}$ and 
$b_{\eta'} e^{i\vartheta_{\eta'}}$ can be obtained from the expressions in Ref.~\cite{FKR}
by making the simple substitution $\mixAng\to \phi_P$. We observe that 
the relations in (\ref{ang-rel}) give a structure of the $B^0_s\to J/\psi \eta'$ amplitude that is
analogous to that for $B^0_s\to J/\psi f_0$ with the tetraquark interpretation of the 
$f_0$. In this case,  there is an additional topology that is specific to the $f_0$ 
tetraquark state. On the other hand, we have an additional contribution to $B^0_s\to J/\psi \eta'$ 
from the gluonic component of the $\eta'$. Using (\ref{ang-rel}), we arrive at 
\begin{equation}\label{A-etap}
{\cal A}_{\eta'} \approx \lambda^2 A \sqrt{\frac{2}{3}}\left[\tilde A^{(c)}_{\rm T} + 
\tilde A^{(ct)}_{\rm P} + 
2 \tilde A^{(c)}_{\rm E} + 2 \tilde A^{(ct)}_{\rm PA}+
\sqrt{\frac{3}{2}}\left(\tilde A_{\rm E,gg}^{(c)} + \tilde A_{\rm PA,gg}^{(ct)}\right)
\tan\phi_G\right]\cos\phi_G
\end{equation}
\begin{equation}\label{b-eta-prime}
b_{\eta'} e^{i\vartheta_{\eta'}} \approx R_b\left[
	\frac{\tilde A^{(ut)}_{\rm P} +\frac{1}{2}\tilde A^{(u)}_{\rm E} + 2\tilde A^{(ut)}_{\rm PA}+
	\sqrt{\frac{3}{2}}\tilde A_{\rm PA, gg}^{(ut)}\tan\phi_G}
	{\tilde A^{(c)}_{\rm T} + \tilde A^{(ct)}_{\rm P} + 2 \tilde A^{(c)}_{\rm E} + 
	2 \tilde A^{(ct)}_{\rm PA}+\sqrt{\frac{3}{2}}\left(\tilde A_{\rm E,gg}^{(c)} + 
	\tilde A_{\rm PA,gg}^{(ct)}\right)\tan\phi_G}\right],
\end{equation}
where $\tilde A^{(q)}_{\rm topology, gg}$ denotes a strong amplitude originating 
from the $|gg\rangle$ admixture. As indicated, the gluonic component can 
only contribute through exchange and penguin annihilation topologies, which are 
expected to be small in comparison to the tree and penguin topologies, respectively \cite{FKR}.
A further suppression comes from $\tan^2\phi_G\sim0.1$. It is interesting to note in
passing that the dynamics are different in $B\to K\eta'$ decays, where a gluonic component 
of the $\eta'$ can contribute in the leading penguin topologies.

\boldmath
\section{The $B^0_s\to J/\psi \eta^{(\prime)}$ Observables}\label{sec:obs}
\unboldmath
The Belle collaboration reported the observation of $B^0_s\to J/\psi\eta$ and
evidence for the $B^0_s\to J/\psi \eta'$ decay in 2009, with the following branching ratio 
measurements \cite{Adachi:2009usa}:
\begin{align}
{\rm BR}(B_s^0 \to J/\psi \eta) &= \left[3.32 \pm 0.87 \,(\mbox{stat.}) {}^{+0.32}_{-0.28} \,(\mbox{syst.}) 
\pm 0.42\,(f_s)\right]\times 10^{-4}\label{belle-Bseta} \\
{\rm BR}(B_s^0 \to J/\psi \eta') &= \left[3.1 \pm 1.2 \,(\mbox{stat.}) {}^{+0.5}_{-0.6} \,(\mbox{syst.}) 
\pm 0.38\,(f_{s})\right]\times 10^{-4}.
\end{align}
Here the latter errors refer to the $B_s$ fragmentation function $f_s$.

Using the $SU(3)_{\rm F}$ flavour symmetry, these branching ratios can be related to
that of $B^0_d\to J/\psi K^0$. Taking factorizable $SU(3)_{\rm F}$-breaking corrections
into account yields
\begin{equation}
\left.\frac{\mbox{BR}(B^0_s\to J/\psi \eta^{(\prime)})}{\mbox{BR}(B^0_d\to J/\psi K^0)}
\right|_{\rm fact.}=
\frac{\tau_{B^0_s}}{\tau_{B^0_d}}
\left[\frac{M_{B^0_s}\Phi^{\eta^{(\prime)}}_s}{M_{B^0_d}\Phi^{K^0}_d}\right]^3
\left[\frac{F_{1}^{B^0_s\eta^{(\prime)}}(M_{J/\psi}^2)}{F_1^{B^0_dK^0}(M_{J/\psi}^2)}\right]^2,
\end{equation}
where the $\tau_{B^0_q}$ and $M_{B^0_q}$ are the $B^0_q$ lifetimes and masses, respectively, 
\begin{equation}
	\Phi_q^{P}  \equiv \sqrt{\left[1-\left(\frac{M_{P} + 
	M_{J/\psi}}{M_{B_q}}\right)^2\right]\left[1-\left(\frac{M_{P}
	- M_{J/\psi}}{M_{B_q}}\right)^2\right]}
\end{equation}
denotes the phase-space factor for $B_q^0 \to J/\psi P$ decays, and the 
$F_1^{B^0_qP}(M^2_{J/\psi})$ are hadronic form factors of quark currents 
(for a detailed discussion, see Ref.~\cite{FKR}). These relations have been 
used previously to predict the $B^0_s\to J/\psi \eta^{(')}$ branching ratios \cite{skands,CFW}.

We advocate to use the measured values to probe non-factorizable $SU(3)_{\rm F}$-breaking 
corrections. To this end we define the quantities
\begin{equation}
K_{SU(3)}^{\eta^{(\prime)}} \equiv
\frac{\tau_{B^0_d}}{\tau_{B^0_s}}
\left[\frac{M_{B^0_d}\Phi^{K^0}_d}{M_{B^0_s}\Phi^{\eta^{(\prime)}}_s}\right]^3
\left[\frac{F_1^{B^0_dK^0}(M_{J/\psi}^2)}{F_{1}^{B^0_s\eta^{(\prime)}}(M_{J/\psi}^2)}\right]^2
\frac{\mbox{BR}(B^0_s\to J/\psi \eta^{(\prime)})}{\mbox{BR}(B^0_d\to J/\psi K^0)},
\end{equation}
where $K_{SU(3)}^{\eta^{(\prime)}}=1$ in the case of vanishing non-factorizable 
$SU(3)_{\rm F}$-breaking corrections. Since non-perturbative calculations of the 
$F_{1}^{B^0_s\eta^{(\prime)}}(M_{J/\psi}^2)$ form factors are not yet available, we 
project out on the $|\eta_s\rangle$ component in (\ref{eta-1}) and write
\begin{eqnarray}
F_1^{B^0_s\eta}(M^2_{J/\psi}) & = & -\sin\phi_P F_1^{B^0_dK^0}(M^2_{J/\psi}) \label{FF-eta-1}\\
F_1^{B^0_s\eta'}(M^2_{J/\psi}) & = & \cos\phi_G\cos\phi_P F_1^{B^0_dK^0}(M^2_{J/\psi}),
\end{eqnarray}
where we neglect $SU(3)_{\rm F}$-breaking corrections originating from the down and 
strange spectator quarks. Using 
$\mbox{BR}(B_d^0\to J/\psi K^0)=(8.71\pm0.32)\times10^{-4}$ \cite{PDG} then yields
\begin{equation}\label{pred-1}
K_{SU(3)}^{\eta}=\left[\frac{\sin 40^\circ}{\sin\phi_P}\right]^2
\times \left( 0.87 \pm 0.27 \right),\quad
K_{SU(3)}^{\eta^{\prime}}=\left[\frac{\cos 20^\circ}{\cos\phi_G}\right]^2
\left[\frac{\cos 40^\circ}{\cos\phi_P}\right]^2 \times \left( 0.8 \pm
0.4 \right).
\end{equation}
These numbers do not indicate any anomalous behaviour, although the currently 
large errors preclude us from drawing stronger conclusions. In Section~\ref{sec:mix},
we will return to the $B^0_s\to J/\psi \eta^{(\prime)}$ branching ratios, using them to
extract the mixing angles $\phi_P$ and $\phi_G$. 

A simple observable that is offered by the $B^0_s\to J/\psi \eta^{(\prime)}$ decays
is their effective lifetime, which is defined through the time expectation value \cite{FK}
\begin{equation}\label{lifetime-def}
 \tau_{J/\psi \eta^{(\prime)}} \equiv \frac{\int^\infty_0 t\ \langle \Gamma(B_s(t)\to J/\psi 
 \eta^{(\prime)})\rangle\ dt} {\int^\infty_0 \langle \Gamma(B_s(t)\to J/\psi \eta^{(\prime)})\rangle\ dt}
\end{equation}
of the untagged rate
\begin{align}\label{eqn:untagged}
	\langle \Gamma(B_s(t)\to J/\psi \eta^{(\prime)})\rangle 
	& \equiv \Gamma(B^0_s(t)\to J/\psi \eta^{(\prime)})+
	\Gamma(\bar B^0_s(t)\to J/\psi \eta^{(\prime)})  \notag\\
	& =  R_{{\rm H}}^{J/\psi \eta^{(\prime)}}\, e^{-\Gamma_{\rm H}^{(s)} t} 
	+ R_{{\rm L}}^{J/\psi \eta^{(\prime)}}\, e^{-\Gamma_{\rm L}^{(s)} t}\\
	\propto &\ e^{-\Gamma_st}\left[ \cosh\left(\frac{\Delta\Gamma_s t}{2}\right)+
	{\cal A}_{\rm \Delta\Gamma}^{J/\psi \eta^{(\prime)}}\,\sinh\left(\frac{\Delta\Gamma_s t}
	{2}\right)\right],\notag
\end{align}
where L and H denote the light and heavy $B_s$ mass eigenstates, respectively, 
$\Delta\Gamma_s \equiv \Gamma^{(s)}_{\rm L} - \Gamma^{(s)}_{\rm H}$ and 
$\Gamma_s \equiv ( \Gamma^{(s)}_{\rm L} + \Gamma^{(s)}_{\rm H})/2 = \tau_{B_s}^{-1}$.
This lifetime is equivalent to that resulting from a fit of the two exponentials 
in \eqref{eqn:untagged} to a single exponential \cite{HM}. 
The effective lifetime can thus be expressed as
\begin{equation}	\label{eqn:lifetime}
	\frac{\tau_{J/\psi  \eta^{(\prime)}}}{\tau_{B_s}}
	=\frac{1}{1-y_s^2} \left[\frac{1+2\,\Adel^{J/\psi  \eta^{(\prime)}} y_s+y_s^2}{1+ 
	\Adel^{J/\psi  \eta^{(\prime)}} y_s}\right],
\end{equation}
where $y_s\equiv \Delta\Gamma_s/(2\Gamma_s)$. The most recent update for the 
theoretical calculation of the $B_s$ width difference is given as follows \cite{NL}:
\begin{equation}\label{DG-SM}
\frac{\Delta\Gamma_s^{\rm Th}}{\Gamma_s}
=0.133\pm0.032.
\end{equation}

In Ref.~\cite{FKR}, the evaluation of ${\cal A}_{\rm \Delta\Gamma}$ has 
been discussed in detail for the $B^0_s\to J/\psi f_0$ channel, which has a CP-odd
final state and offers an interesting probe of CP violation \cite{SZ}. 
Since the $\eta^{(\prime)}$ are pseudoscalar mesons with quantum numbers 
$J^{PC}=0^{-+}$, the final states of $B^0_{s,d}\to J/\psi \eta^{(\prime)}$ are CP even.
This sign difference results in 
\begin{equation}\label{ADG}
{\cal A}_{\rm \Delta\Gamma}^{J/\psi \eta^{(\prime)}}=
-\sqrt{1-C_{J/\psi \eta^{(\prime)}}^2}\cos(\phi_s+\Delta\phi_{J/\psi  \eta^{(\prime)}}).
\end{equation}
Here $C_{J/\psi \eta^{(\prime)}}$ describes direct CP violation, whereas 
\begin{equation}\label{phis}
\phi_s\equiv\phi_s^{\rm SM}+\phi_s^{\rm NP}
\end{equation}
denotes the $B_s^0$--${\bar B}_s^0$ mixing phase, where \cite{Charles:2011va}
\begin{equation}\label{SM-phiS}
\phi_s^{\rm SM}\equiv-2\beta_s=-(2.08\pm0.09)^\circ
\end{equation}
and $\phi_s^{\rm NP}$ are the SM and NP pieces, respectively.
The quantity $\Delta\phi_{J/\psi  \eta^{(\prime)}}$ is a hadronic phase shift, which 
can be obtained from 
\begin{equation}\label{tDelphi}
\tan\Delta\phi_{J/\psi  \eta^{(\prime)}}=
\frac{2 \, \epsilon \, b_{\eta^{(\prime)}}\cos\vartheta_{\eta^{(\prime)}} \sin\gamma+
\epsilon^2 \, b_{\eta^{(\prime)}}^2\sin2\gamma}{1+ 2 \, \epsilon \, b_{\eta^{(\prime)}}
\cos\vartheta_{\eta^{(\prime)}}\cos\gamma+\epsilon^2 \, b_{\eta^{(\prime)}}^2\cos2\gamma}.
\end{equation}

\begin{figure}[t]
   \centering
   \begin{tabular}{cc}
   	  \includegraphics[width=7.9truecm]{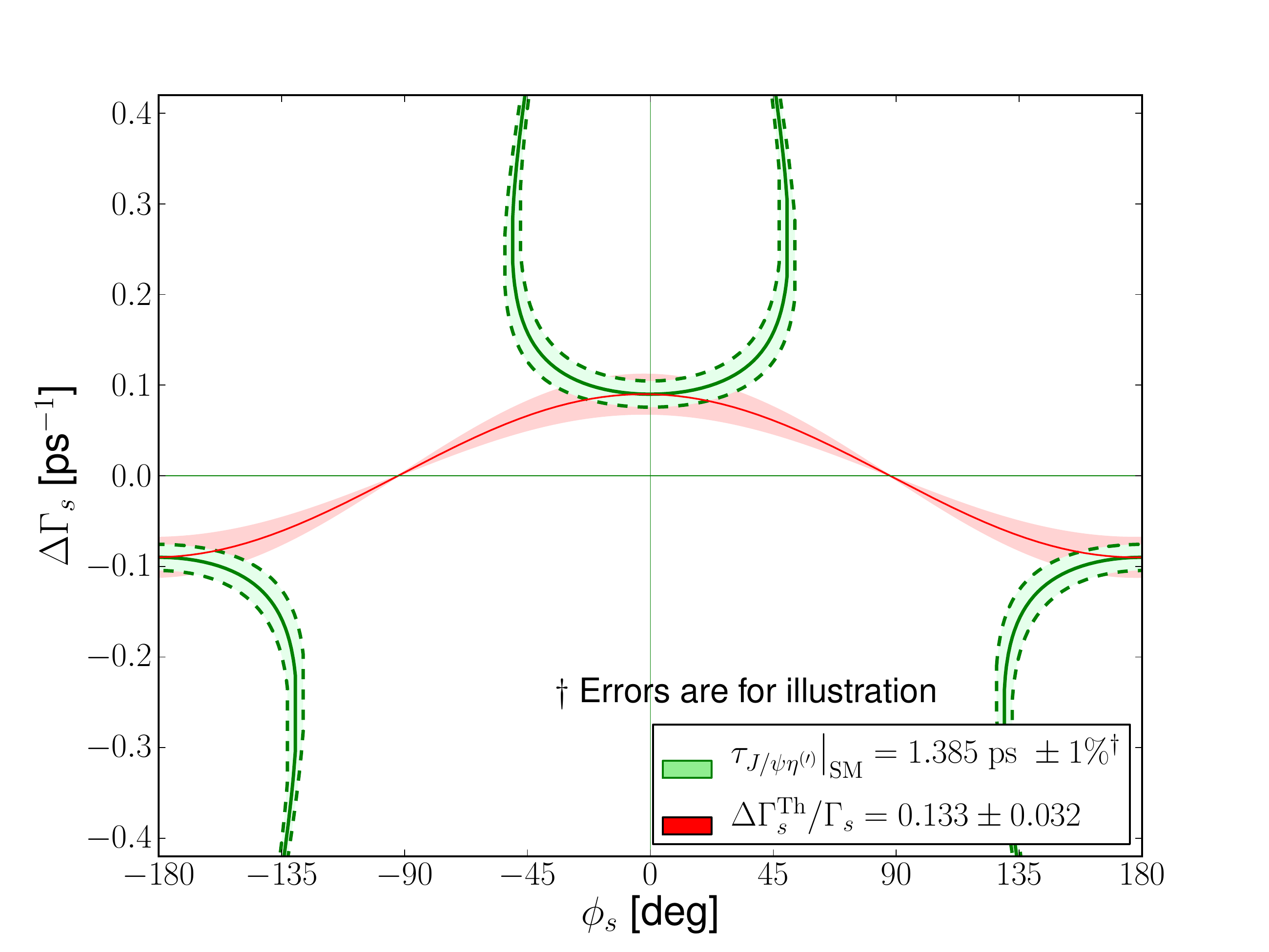} &
      \includegraphics[width=7.9truecm]{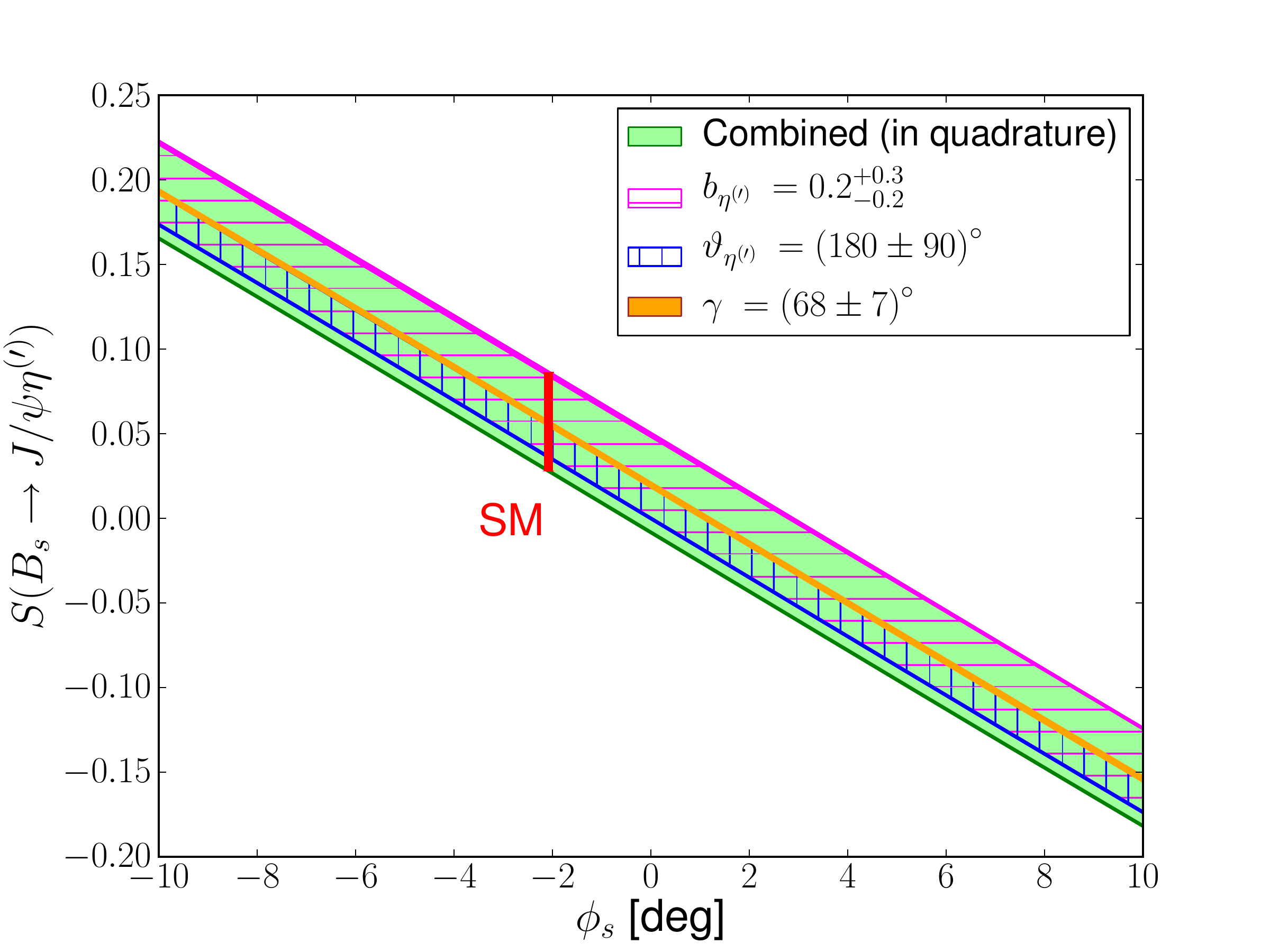}
   \end{tabular}
   \caption{{\it Left panel:} the $B^0_s\to J/\psi \eta^{(\prime)}$ effective lifetime as a constraint 
   in the $\phi_s$--$\Delta\Gamma_s$ plane. For illustration we have chosen a central value 
   compatible with the SM values of $\phi_s$ and $\Delta\Gamma_s$ given in 
   \eqref{SM-phiS} and \eqref{ys}, respectively, and have assumed a lifetime measurement
   with 1\% uncertainty. {\it Right panel}: the mixing-induced CP asymmetry of 
   $B^0_s\to J/\psi \eta^{(\prime)}$ as a function of $\phi_s$, assuming $\gamma=(68\pm7)^\circ$, 
   $0\leq b_{\eta^{(\prime)}} \leq 0.5$ and $90^\circ\leq \vartheta_{\eta^{(\prime)}}\leq270^\circ$  
   for the calculation of the error band. We show only the region close to the SM case.}	
	\label{fig:lifetime}
\end{figure}

We observe that the hadronic parameters, which are poorly known, enter 
$\Delta\phi_{J/\psi\eta^{(\prime)}}$ in a doubly Cabibbo-suppressed way. 
Therefore the effective lifetimes 
turn out to be very robust with respect to the hadronic corrections, in analogy to the 
situation in $B^0_s\to J/\psi f_0$. As in Ref.~\cite{FKR}, we use 
$\gamma=(68\pm7)^\circ$ in order to illustrate the hadronic effects. As far as the 
hadronic parameters are concerned, we consider the ranges
\begin{equation}\label{b-range}
0\leq b_{\eta^{(\prime)}}\leq 0.5, \quad 90^\circ \leq\vartheta_{\eta^{(\prime)}}\leq 270^\circ.
\end{equation} 
Due to the factor $R_b\sim0.5$ in (\ref{b-eta}) and (\ref{b-eta-prime}) the range 
for $b_{\eta^{(\prime)}}$ is conservative. The range for the strong phase is motivated 
by the topological structure of (\ref{b-eta}) and (\ref{b-eta-prime}). It is also supported by 
an $SU(3)_{\rm F}$ analysis of $B^0_d\to J/\psi \pi^0$ data, where the counterparts of 
$b_{\eta^{(\prime)}}$ and $\vartheta_{\eta^{(\prime)}}$ in $B^0_d\to J/\psi K^0$, $a$ and 
$\theta$, are found to be $a\in[0.15,0.67]$ and $\theta\in[174,212]^\circ$ at the $1\,\sigma$ 
level~\cite{FFJM} (see also Ref.~\cite{CPS}). For the 
ranges given in \eqref{b-range} the hadronic phase shift takes values in the interval
\begin{equation}
	\Delta\phi_{J/\psi \eta^{(\prime)}}	\in [-3^\circ, 0^\circ ].
\end{equation}
Likewise, the direct CP asymmetry satisfies $|C_{J/\psi \eta^{(\prime)}}|\lsim0.05$ under 
these assumptions and thereby has a negligible impact on 
${\cal A}_{\rm \Delta\Gamma}^{J/\psi \eta^{(\prime)}}$. 

Once the $B^0_s\to J/\psi \eta^{(\prime)}$ effective lifetimes have been measured they can be 
converted into contours in the $\phi_s$--$\Delta\Gamma_s$ plane, as was pointed out in 
Ref.~\cite{FK-lifetime}.
The interesting feature of this analysis is that it does not rely on the theoretical value 
$\Delta\Gamma^{\rm Th}_s$, in contrast to the lifetime analysis given in Ref.~\cite{FKR}. 
Furthermore, using complementary information from a second CP-odd final state, such 
as $B^0_s\to J/\psi f_0$, the mixing parameters $\phi_s$ and $\Delta\Gamma_s$ can be extracted.
These can then be compared with information from other analyses, such as $B^0_s\to J/\psi \phi$. 
In the left panel of Fig.~\ref{fig:lifetime}, we show for illustration the lifetime contour that is 
compatible with the values of $\phi_s$ and $\Delta\Gamma_s$ given in \eqref{SM-phiS} 
and \eqref{DG-SM}, respectively. The corresponding theoretical SM prediction for the effective 
lifetimes is
\begin{equation}
	\left. \tau_{J/\psi \eta^{(\prime)}}\right|_{\rm SM} = \left(1.385\pm 0.029 \right)\ {\rm ps},
\end{equation}
where we have used $\tau_{B_s}=(1.477^{+0.021}_{-0.022})$\,ps \cite{HFAG}.
In the same plot we have also included a contour that arises from the plausible 
assumption that NP affects $\Delta\Gamma_s$ only through $B^0_s$--$\bar B^0_s$ mixing
\cite{Grossman}, implying the relation
\begin{equation}\label{ys}
y_s=\frac{\Delta\Gamma_s^{\rm Th}\cos\tilde\phi_s}{2\Gamma_s}=
y_s^{\rm Th}\cos (\tilde\phi_s^{\rm SM}+\phi_s^{\rm NP}).
\end{equation}
Here $\phi_s^{\rm NP}$ is the NP $B^0_s$--$\bar B^0_s$ mixing phase,
which also enters the phase $\phi_s$ defined in (\ref{phis}) on which 
$\Adel^{J/\psi \eta^{(\prime)}}$ depends, whereas the SM piece is given by  
$\tilde\phi_s^{\rm SM}=(0.22\pm0.06)^\circ$ \cite{NL}.

There is an interesting trend of the current Tevatron and LHCb data to favour a value of 
$\Delta\Gamma_s$ larger than (\ref{DG-SM}), which raises the question of whether the 
corresponding theoretical analysis of $\Delta\Gamma_s^{\rm Th}$ fully includes all hadronic 
long-distance contributions \cite{FK-lifetime}. It will be interesting to see if this trend will 
persist with future data or if it will eventually disappear. 

A tagged analysis of $B^0_s\to J/\psi \eta^{(\prime)}$ decays allows the measurement
of the following CP-violating rate asymmetry:
\begin{equation}\label{t-dep-asym}
\frac{\Gamma(B_s(t)\to J/\psi \eta^{(\prime)})-\Gamma(\bar B_s(t)\to J/\psi 
\eta^{(\prime)})}{\Gamma(B_s(t)\to J/\psi \eta^{(\prime)})+
\Gamma(\bar B_s(t)\to J/\psi \eta^{(\prime)})}
=\frac{C_{J/\psi \eta^{(\prime)}}\cos(\Delta M_st) -
S_{J/\psi \eta^{(\prime)}} \sin(\Delta M_st)}{\cosh(\Delta\Gamma_st/2)+
{\cal A}^{_{J/\psi \eta^{(\prime)}}}_{\Delta\Gamma}\sinh(\Delta\Gamma_st/2)} ,
\end{equation}
where $C_{J/\psi \eta^{(\prime)}}$ is the direct CP asymmetry that we have already
encountered in (\ref{ADG}), and
\begin{equation}
S_{J/\psi \eta^{(\prime)}}=-\sqrt{1-C_{J/\psi \eta^{(\prime)}}^2}
\sin(\phi_s+\Delta\phi_{J/\psi  \eta^{(\prime)}})
\end{equation}
describes mixing-induced CP violation. The minus sign differs from the mixing-induced
CP asymmetry of the $B^0_s\to J/\psi f_0$ channel \cite{FKR} 
because of the opposite CP eigenvalues
of the final states. In the right panel of Fig.~\ref{fig:lifetime}, we show the dependence of 
$S_{J/\psi \eta^{(\prime)}}$ on the mixing phase $\phi_s$ and illustrate 
how the hadronic SM uncertainties as well as 
the uncertainties on $\gamma$ propagate through. We observe that a future measurement of 
the mixing-induced CP asymmetry in the range 
\begin{equation}
0.03\lsim S_{J/\psi \eta^{(\prime)}}\lsim 0.09
\end{equation} 
would not allow us to distinguish the SM from CP-violating NP contributions to 
$B^0_s$--$\bar B^0_s$ mixing. Should we encounter such a situation, more information
would be required to accomplish this task. In this respect, things are similar to analyses
of CP violation in $B^0_s\to J/\psi f_0$ \cite{FKR} and $B^0_s\to J/\psi \phi$ \cite{FFM}.
In the case of the $B^0_s\to J/\psi \eta^{(\prime)}$ decays, the hadronic uncertainties
can be controlled with the help of the $B^0_d\to J/\psi \eta^{(\prime)}$ channels.

\boldmath
\section{The $B^0_d\to J/\psi \eta^{(\prime)}$ Control Channels}\label{sec:contr}
\unboldmath
The leading contributions to the $B^0_d\to J/\psi \eta$ decay originate from 
$\bar b\to \bar c c\bar d$ quark-level processes. It is the formal counterpart of the
$B^0_d\to J/\psi f_0$ mode discussed in Ref.~\cite{FKR}. Following 
this discussion, we write
\begin{equation}\label{ampl-1-p}
A(B^0_d\to J/\psi \eta)=-\lambda{\cal A}'_\eta 
\left [1- b'_\eta e^{i\vartheta'_\eta} e^{i\gamma}  \right]
\end{equation}
with
\begin{equation}
{\cal A}'_\eta = -\lambda^2A\left[  
\sin\phi_P\left\{ \tilde{A}^{(c)}_{\rm E} + \tilde{A}^{(ct)}_{\rm PA}\right\} -
	\frac{1}{\sqrt{2}}\cos\phi_P \left\{ \tilde{A}^{(c)}_{\rm T} + \tilde{A}^{(ct)}_{\rm P} + 2 \tilde{A}^
	{(c)}_{\rm E} +	2 \tilde{A}^{(ct)}_{\rm PA}\right\} \right]
\end{equation}
and
\begin{equation}
b'_\eta e^{i\vartheta'_\eta} = R_b\left[
	\frac{\sin\phi_P\left\{ \tilde{A}^{(ut)}_{\rm PA}\right\} -
	\frac{1}{\sqrt{2}}\cos\phi_P \left\{\tilde{A}^{(ut)}_{\rm P} + \tilde{A}^{(u)}_{\rm E} +
	2 \tilde{A}^{(ut)}_{\rm PA}\right\}}
	{\sin\phi_P\left\{ \tilde{A}^{(c)}_{\rm E} + \tilde{A}^{(ct)}_{\rm PA}\right\} -
	\frac{1}{\sqrt{2}}\cos\phi_P \left\{ \tilde{A}^{(c)}_{\rm T} + \tilde{A}^{(ct)}_{\rm P} + 2 \tilde{A}^
	{(c)}_{\rm E} +	2 \tilde{A}^{(ct)}_{\rm PA}\right\}}\right],
	\label{bthetaPrime}
\end{equation}
where we have used $SU(3)_{\rm F}$ arguments to identify the topological amplitudes
with those in (\ref{A-expr}) and (\ref{b-expr}). The simplified expressions in (\ref{ang-rel}) 
yield
\begin{equation}\label{Apeta}
{\cal A}'_\eta \approx \lambda^2A\sqrt{\frac{1}{3}}\left[ 
\tilde A^{(c)}_{\rm T} + \tilde A^{(ct)}_{\rm P} + \tilde A^{(c)}_{\rm E} + \tilde A^{(ct)}_{\rm PA} \right],
\end{equation}
\begin{equation}
b'_{\eta}  e^{i\vartheta'_{\eta}} \approx R_b\left[
	\frac{\tilde A^{(ut)}_{\rm P} + \tilde A^{(u)}_{\rm E} + \tilde A^{(ut)}_{\rm PA}}
	{\tilde A^{(c)}_{\rm T} + \tilde A^{(ct)}_{\rm P} + \tilde A^{(c)}_{\rm E} + 
	\tilde A^{(ct)}_{\rm PA}}\right].\label{bPrime-eta}
\end{equation}
The key difference of the $B^0_d\to J/\psi \eta$ decay with respect to its $B^0_s\to J/\psi \eta$
counterpart is that the hadronic parameter $b'_\eta e^{i\vartheta'_\eta}$ does not enter 
(\ref{ampl-1-p}) in a doubly Cabibbo-suppressed way. Consequently, its impact is ``magnified" 
in the $B^0_d\to J/\psi \eta$ observables. On the other hand, the branching ratio does suffer 
from a $\lambda^2$ suppression.\footnote{Analogous features apply to 
$B^0_s\to J/\psi K_{\rm S}$ \cite{RF-BpsiK,DeBFK}, $B^0_d\to J/\psi f_0$ 
\cite{FKR}, and $B^0_d\to J/\psi \pi^0$ \cite{FFJM,CPS}.}

As discussed in detail in Ref.~\cite{FKR}, the exchange and penguin amplitudes play a 
minor role and can be probed through the $B^0_d\to J/\psi\phi$ and $B^0_s\to J/\psi\pi^0$ 
decays, where already the currently available upper bound on the branching ratio of the
former decay allows us to put the upper bound
\begin{equation}
\left |\frac{\tilde A^{(c)}_{\rm E} + \tilde A^{(ct)}_{\rm PA}}{\tilde A^{(c)}_{\rm T}}\right| \lsim 0.1. 
\end{equation}
Neglecting these contributions and using the $SU(3)_{\rm F}$ symmetry (as
we have already implicitly done in the expression given above), we obtain 
\begin{equation}\label{b-rel}
b_{\eta} e^{i\vartheta_{\eta}}\stackrel{SU(3)_{\rm F}}{=} R_b\left[
	\frac{\tilde A^{(ut)}_{\rm P}}
	{\tilde A^{(c)}_{\rm T} + \tilde A^{(ct)}_{\rm P}}\right]\stackrel{SU(3)_{\rm F}}{=}
	b'_{\eta} e^{i\vartheta'_{\eta}}.
\end{equation}
Interestingly, the dependence on $\phi_P$ drops out if the exchange and penguin 
annihilation contributions are neglected. Since the parameters 
$b'_{\eta}$ and $\vartheta'_{\eta}$ can be determined  from the $B^0_d\to J/\psi \eta$ 
observables in a clean way (in analogy to the discussion for $B^0_d\to J/\psi f_0$ in 
Ref.~\cite{FKR}), we can control the penguin effects in the $B^0_s\to J/\psi \eta$ 
observables. 

As the $b_{\eta}^{(\prime)} e^{i\vartheta_{\eta}^{(\prime)}}$ are ratios of hadronic amplitudes,  
we expect (\ref{b-rel}) to be robust with respect to $SU(3)_{\rm F}$-breaking corrections. 
Should the $B^0_d\to J/\psi \eta$ data favour a small value of $b'_\eta$, the exchange and 
penguin annihilation amplitudes could contribute significant uncertainties in relating 
$b'_\eta$ to $b_\eta$. 
However, the doubly Cabibbo-suppressed corrections to the mixing-induced CP violation in 
$B^0_s\to J/\psi \eta$ would then be tiny anyway.

\begin{figure}[t]
   \centering
   \begin{tabular}{cc}
	   \includegraphics[width=7.9truecm]{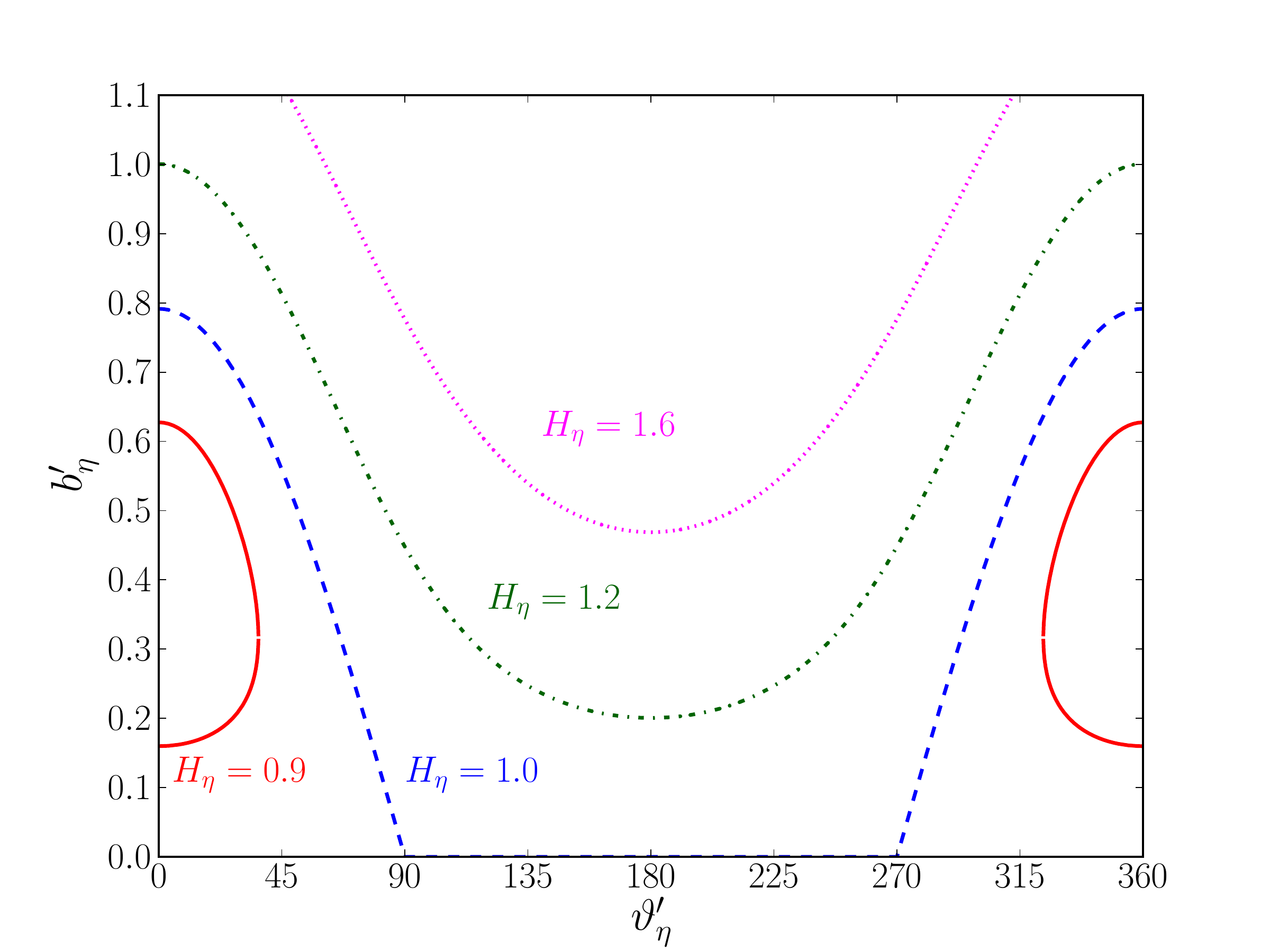} &
	   \includegraphics[width=7.9truecm]{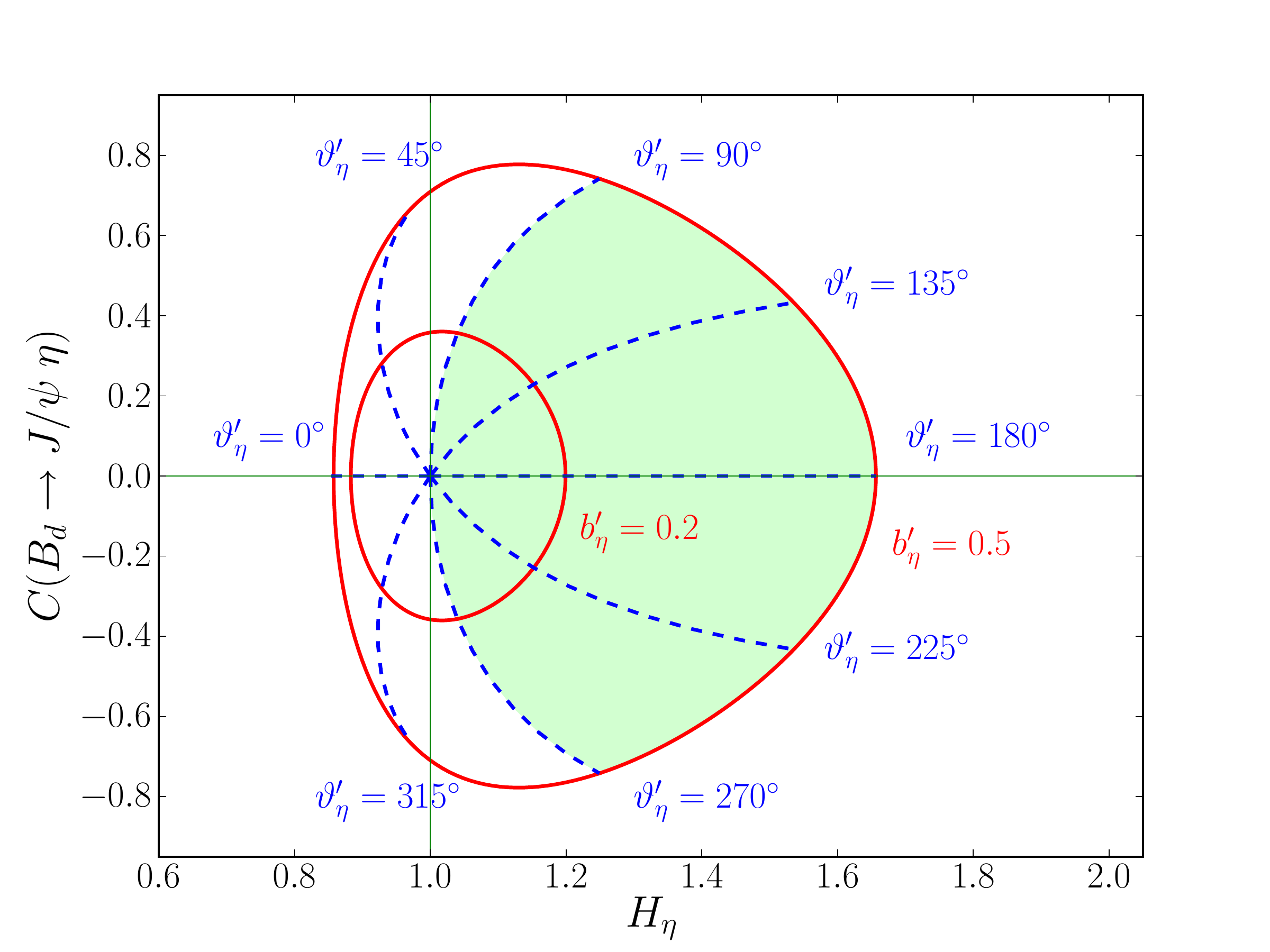}
   \end{tabular}
   \caption{ {\it Left panel:} constraints in the $\vartheta'_{\eta}$--$b'_{\eta}$ plane for various 
   values of $H_\eta$.  {\it Right panel:} correlation between $H_\eta$ 
   and the direct CP asymmetry  of $B^0_d\to J/\psi \eta$, where the solid rings correspond 
   to $b'_{\eta}=0.2$ and 0.5 with $\vartheta'_{\eta}$ allowed to vary; likewise, the dashed 
   lines are fixed points of $\vartheta'_{\eta}$ with $b'_{\eta}$ allowed to vary. In both plots,
   we have assumed $\gamma=68^\circ$.}\label{fig:H}
\end{figure}

The $B^0_d\to J/\psi \eta$ decay was observed by the Belle collaboration \cite{Belle-Jpsi-eta},
with
\begin{equation}\label{BdJpsieta-Belle}
\mbox{BR}(B^0_d\to J/\psi \eta)=[9.5\pm1.7 ({\rm stat.})\pm0.8({\rm syst.})]\times10^{-6},
\end{equation}
which is consistent with the estimates given in Ref.~\cite{skands}. We can use 
this measurement to obtain a first constraint on the hadronic parameters with the help of 
\begin{equation}\label{H-def}
	H_{\eta} \equiv \frac{1}{\epsilon}
	\left|\frac{{\cal A}_\eta}{{\cal A}'_\eta}\right|^2
	\left(\frac{M_{B^0_s}\Phi^{\eta}_s}{M_{B^0_d}\Phi^{\eta}_d}\right)^3
	\frac{\tau_{B^0_s}}{\tau_{B^0_d}}\,
	\frac{\mbox{BR}(B^0_d\to J/\psi \eta)}{\mbox{BR}(B^0_s\to J/\psi \eta)},
\end{equation}
where the branching ratios refer to CP-averaged combinations. The formulae given above
yield the following expression in terms of $\gamma$ and the hadronic parameters:
\begin{equation}\label{H-expr-3}
H_{\eta}=\frac{1-2b'_\eta\cos\vartheta'_\eta\cos\gamma+b_\eta^{\prime2}}{1+2\epsilon b_\eta 
\cos\vartheta_\eta\cos\gamma+\epsilon^2b_\eta^2}.
\end{equation}
In order to extract $H_{\eta}$ from the branching ratios, we have to calculate the $SU(3)$-breaking
ratio of the ${\cal A}_\eta$ and ${\cal A}'_\eta$ amplitudes. Using the factorization approximation
and keeping only the leading tree contributions gives
\begin{equation}\label{A-rat}
\left|\frac{{\cal A}_\eta}{{\cal A}'_\eta}\right|_{\rm fact.}= -\sqrt{2}\tan\phi_P
\left[ \frac{F_1^{B^0_d K^0}(M_{J/\psi}^2)}{F_1^{B^0_d\pi^-}(M_{J/\psi}^2)}\right],
\end{equation}
where we have -- as in (\ref{FF-eta-1}) --  also neglected $SU(3)_{\rm F}$-breaking corrections
that originate from  the down and strange spectator quarks.
Using the form factors
\begin{equation}
F_1^{B^0_dK^0}(M_{J/\psi}^2) = 0.615\pm{0.076}, \quad
F_1^{B^0_d\pi^-}(M_{J/\psi}^2)=0.49 \pm 0.06
\end{equation}
calculated with the leading-order light-cone QCD sum-rule results of 
Ref.~\cite{Ball:2004ye}, as well as the measured $B^0_{s,d}\to J/\psi \eta$ 
branching ratios given earlier, we finally arrive at 
\begin{equation}\label{H-det}
H_\eta\times \left[\frac{\tan 40^\circ}{\tan\phi_P}\right]^2
=  1.28
{}^{+0.61}_{-0.39}\big|_{\rm BR}
{}^{+0.50}_{-0.40}\big|_{\rm FF}
=  1.28^{+0.79}_{-0.56}.
\end{equation}
The errors reflect only the experimental and form-factor uncertainties 
and do not take non-factorizable $SU(3)$-breaking corrections into account.
Using the factorization tests in \eqref{pred-1}, a future more precise measurement
of the $B^0_s\to J/\psi \eta$ branching ratio should give us better quantitative 
insights into these effects.\footnote{Recent studies of other $B_{(s)}$-meson 
decays indicate small effects of this kind \cite{FK,FST-fact}.}   In Fig.~\ref{fig:H}, we convert this 
result into contours in the $\vartheta'_\eta$--$b'_\eta$ plane (see Ref.~\cite{FKR} for details).

As soon as measurements of the CP asymmetries for $B^0_d\to J/\psi \eta$ 
become available we will be able to determine $b'_\eta$ and $\theta'_\eta$ in a clean way.
Subsequently, we can determine $H_\eta$ through (\ref{H-expr-3}). Using then the 
information from the branching ratios and (\ref{H-def}) and (\ref{A-rat}), we can determine 
$|\tan\phi_P|$. Alternatively, assuming that we will have a sharp picture of $\phi_P$
by the time these measurements become available (see also Section~\ref{sec:mix}), 
we can perform another test of non-factorizable $SU(3)_{\rm F}$-breaking corrections.

The counterparts of the hadronic parameters in (\ref{A-etap}) and (\ref{b-eta-prime}) for
$B^0_d\to J/\psi\eta'$  are
\begin{equation}\label{APrime-eta-prime}
{\cal A'}_{\eta'} \approx
\lambda^2 A \sqrt{\frac{1}{6}}\left[\tilde A^{(c)}_{\rm T} + \tilde A^{(ct)}_{\rm P} + 
4 \tilde A^{(c)}_{\rm E} + 4 \tilde A^{(ct)}_{\rm PA}+
\sqrt{6}\left(\tilde A_{\rm E,gg}^{(c)} + \tilde A_{\rm PA,gg}^{(ct)}\right)\tan\phi_G\right]
\cos\phi_G
\end{equation}
and
\begin{equation}\label{bPrime-eta-prime}
b'_{\eta'} e^{i\vartheta'_{\eta'}} \approx R_b\left[
	\frac{\tilde A^{(ut)}_{\rm P} + \tilde A^{(u)}_{\rm E} + 4 \tilde A^{(ut)}_{\rm PA}+
	\sqrt{6}\tilde A_{\rm PA,gg}^{(ut)}\tan\phi_G}{\tilde A^{(c)}_{\rm T} + \tilde A^{(ct)}_{\rm P} + 
	4 \tilde A^{(c)}_{\rm E} + 4 \tilde A^{(ct)}_{\rm PA}+
	\sqrt{6}\left(\tilde A_{\rm E,gg}^{(c)} + \tilde A_{\rm PA,gg}^{(ct)}\right)
	\tan\phi_G}\right],
\end{equation}
respectively. 
Neglecting the exchange and penguin annihilation topologies, the control of the hadronic 
parameters in the $B^0_s\to J/\psi \eta'$ observables by means of the $B^0_d\to J/\psi \eta'$ 
mode is analogous to the case of  the $B^0_{s,d}\to J/\psi \eta$ channels.

\boldmath
\section{Determination of the $\eta$--$\eta'$ Mixing Parameters}\label{sec:mix}
\unboldmath
Let us finally discuss determinations of the $\eta$--$\eta'$ mixing parameters through
measurements of the $B^0_{s,d}\to J/\psi \eta^{(\prime)}$ branching ratios.
If we project out on the singlet states in \eqref{eta-1} and \eqref{eta-2} and assume that
the exchange and penguin annihilation topologies give negligible contributions, we 
obtain the relation
\begin{equation}\label{trig}
R_s\equiv\frac{\mbox{BR}(B^0_s\to J/\psi\eta')}{\mbox{BR}(B^0_s\to J/\psi\eta)}
\left(\frac{\Phi_s^\eta}{\Phi_s^{\eta'}}\right)^3=
\frac{\cos^2\phi_G}{\tan^2\phi_P} = 1.3^{+1.5}_{-0.5},
\end{equation}
which does not assume $SU(3)_{\rm F}$ or factorization; the numerical value 
corresponds to the Belle result \cite{Adachi:2009usa}. The same expression with 
$\phi_G=0$ has already been given in Ref.~\cite{Datta:2001ir}.

In analogy to (\ref{trig}), we introduce the following ratio for the $B_d$ decays:
\begin{equation}\label{BR-e}
R_d\equiv\frac{\mbox{BR}(B^0_d\to J/\psi \eta')}{\mbox{BR}(B^0_d\to J/\psi \eta)}
\left(\frac{\Phi_d^\eta}{\Phi_d^{\eta'}}\right)^3= \cos^2\phi_G \, \tan^2\phi_P.
\end{equation}
Using the experimental value in (\ref{BdJpsieta-Belle}), this expression
results in the prediction
\begin{equation}
\mbox{BR}(B^0_d\to J/\psi \eta')=
\left[\frac{\cos\phi_G}{\cos20^\circ}\right]^2 \left[\frac{\tan\phi_P}{\tan 40^\circ}\right]^2
\times\left(4.7\pm 0.9 \right)\times 10^{-6}.
\end{equation} 
Only the upper bound $\mbox{BR}(B^0_d\to J/\psi \eta')<6.3\times 10^{-5}$ (90\% C.L.) 
is currently available from the BaBar collaboration \cite{Babar-Jpsi-etap}.

\begin{figure}[t]
   \centering
   \begin{tabular}{cc}
   	  \includegraphics[width=7.9truecm]{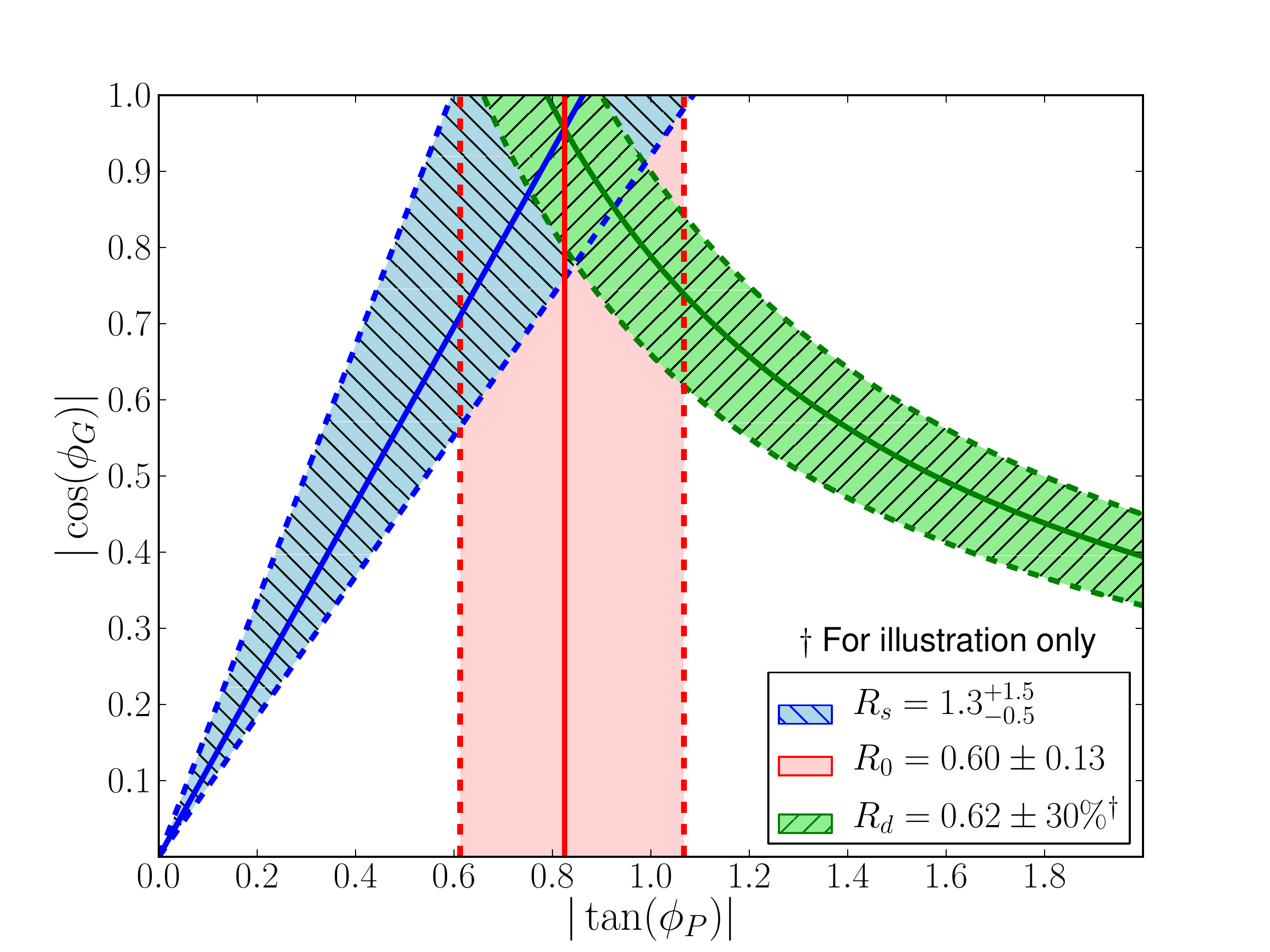} &
      \includegraphics[width=7.9truecm]{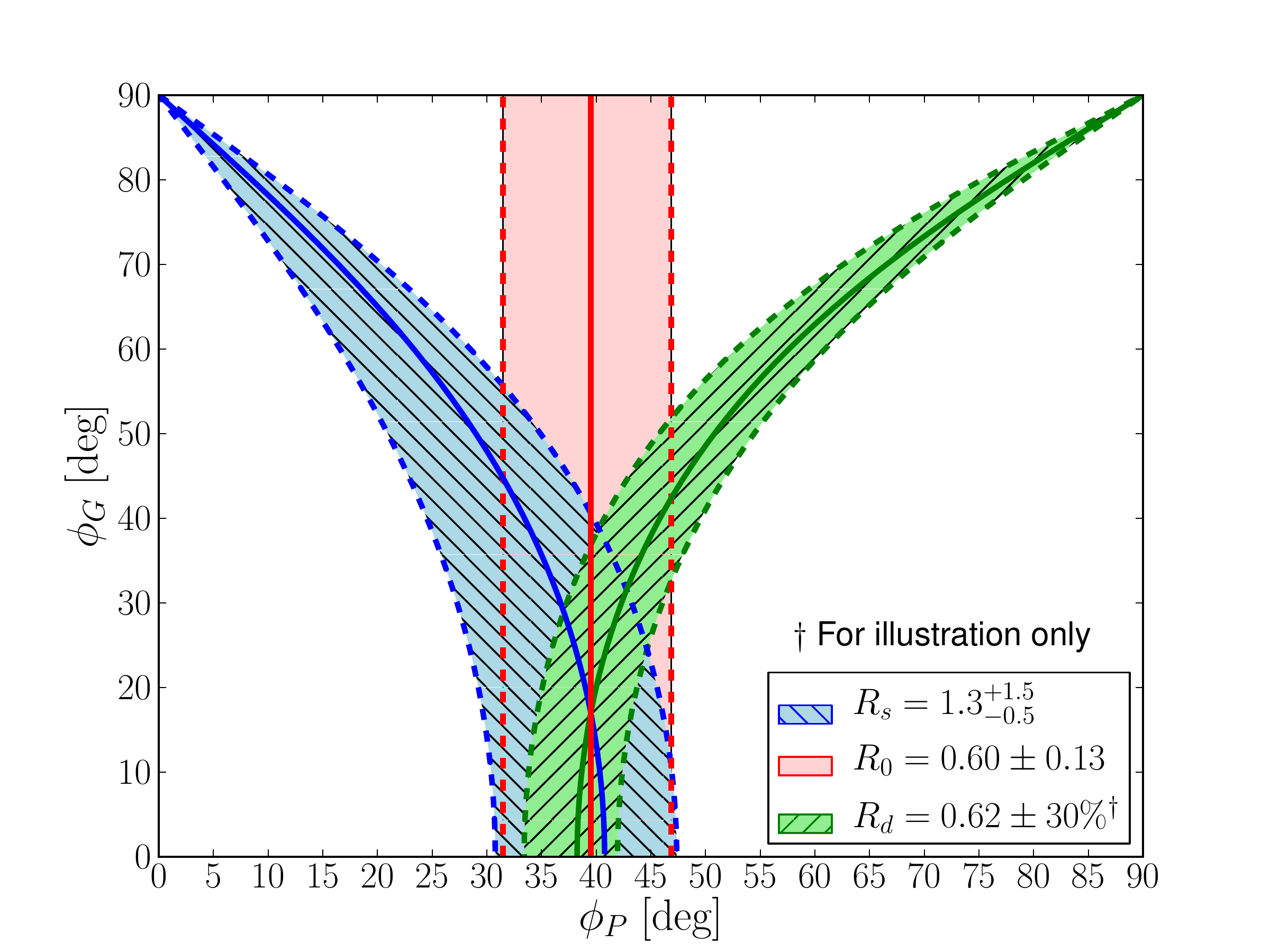}
   \end{tabular}
   \caption{Constraints on the $\eta$--$\eta'$ mixing parameters from
   the $B^0_{s,d}\to J/\psi \eta^{(\prime)}$ and $B^0_d\to J/\psi \pi^0$ branching 
   ratios as discussed in the text. Note that the right panel does not show all the discrete 
   angular ambiguities.
   }\label{fig:mixAngles}
\end{figure}

Once the $B^0_d\to J/\psi \eta^{\prime}$ branching ratio has been measured,  we can use 
\begin{equation}\label{RsRd}
\frac{R_d}{R_s}=\tan^4\phi_P, \quad R_s R_d =\cos^4\phi_G
\end{equation}
to determine the mixing angles up to fourfold discrete ambiguities. It is interesting to note that
the 4th powers in these expressions result in precise determinations of $|\tan\phi_P|$
and $|\cos\phi_G|$ even for branching ratio measurements with significant errors. If we 
assume, for illustration, future measurements of $R_s=1.3\pm0.4$ and $R_d=0.6\pm0.2$,
i.e.\ with precisions of 30\%,  we would obtain $\phi_P =  ( 39.5 \pm 3.1)^\circ$ and 
$|\phi_G|\in[0^\circ,33^\circ ]$.

In Fig.~\ref{fig:mixAngles}, we have illustrated this method, showing the contours 
for the current experimental value of $R_s$ 
in (\ref{trig}) and our illustrative value of $R_d=0.6\pm0.2$. It is interesting to 
include also the constraint from the following ratio \cite{Thomas:2007uy}:
\begin{equation}\label{R0-def}
R_0\equiv\frac{\mbox{BR}(B^0_d\to J/\psi\eta)}{\mbox{BR}(B^0_d\to J/\psi\pi^0)}
\left(\frac{\Phi_d^{\pi^0}}{\Phi_d^{\eta}}\right)^3= \cos^2\phi_P.
\end{equation}
Here penguin annihilation and exchange topologies were again neglected. The penguin 
parameters of the $B^0_d\to J/\psi \pi^0$ decay \cite{FFJM,CPS} are then the
same as in the $B^0_d\to J/\psi \eta^{(\prime)}$ modes. In particular, we expect also the 
same direct and mixing-induced CP asymmetries. 
As (\ref{R0-def}) does not depend on $\phi_G$, we can straightforwardly convert the 
Belle result in (\ref{BdJpsieta-Belle}) with 
BR$(B_d^0 \to J/\psi \pi^0)= (1.76\pm 0.16)\times 10^{-5}$~\cite{PDG} into 
\begin{equation}
\left.\phi_P\right|_{R_0} = \left(40^{+7}_{-8}\right)^\circ.
\end{equation}
The intersection of the corresponding band in Fig.~\ref{fig:mixAngles}  with the
$R_s$ contour  gives $\phi_G = 17^\circ$ for the central values. These results are 
in good agreement with those discussed at the beginning of Section~\ref{sec:ampl}.

\boldmath
\section{Conclusions}\label{sec:concl}
\unboldmath
The $B^0_s\to J/\psi \eta^{(\prime)}$ decays offer interesting insights into the 
$B^0_s$--$\bar B^0_s$ mixing parameters through their effective lifetimes and mixing-induced
CP asymmetries. We have performed an analysis of these observables, focusing on hadronic 
SM corrections which enter in a doubly Cabibbo-suppressed way. It turns out
that the effective lifetimes are particularly robust with respect to these effects. Once they have 
been measured, we can convert the corresponding experimental results into contours in
the $\phi_s$--$\Delta\Gamma_s$ plane. As far as the mixing-induced CP asymmetries are
concerned, measured values within the range $0.03 \lsim S_{J/\psi \eta^{(\prime)}}\lsim 0.09$
would not allow us to distinguish CP-violating NP contributions to $B^0_s$--$\bar B^0_s$ 
mixing from SM effects, unless we can control the hadronic SM corrections. 

We have shown that this can be accomplished with the help of the $B^0_d\to J/\psi \eta^{(\prime)}$
channels and the $SU(3)_{\rm F}$ flavour symmetry. In these decays, the relevant 
hadronic parameters are not doubly Cabibbo-suppressed. Only a 
branching ratio measurement for $B^0_d\to J/\psi \eta$ is available from the Belle 
collaboration, which we have used to obtain the first value of the $H_\eta$ observable.
This observable implies (still pretty poor) constraints for the hadronic parameters. The next 
important step to constrain them in a more stringent way would be the measurement 
of direct CP violation in $B^0_d\to J/\psi \eta$. Other interesting control channels in
this respect are $B^0_s\to J/\psi K_{\rm S}$ and $B^0_d\to J/\psi \pi^0$. If exchange
and penguin annihilation topologies are neglected, they depend on the same penguin
parameters $b'_{\eta^{(\prime)}}$ and $\theta'_{\eta^{(\prime)}}$. It is important to obtain
stronger experimental constraints on these topologies in the future through decays
such as $B^0_d\to J/\psi \phi$ and $B^0_s\to J/\psi \pi^0$.

In addition to exploring CP violation, the $B^0_{s}\to J/\psi \eta^{(\prime)}$ and 
$B^0_{d}\to J/\psi \eta^{(\prime)}$ decays allow us to probe non-factorizable 
$SU(3)_{\rm F}$-breaking effects and offer interesting strategies for determining the 
$\eta$--$\eta'$ mixing parameters from their ratios of branching ratios, $R_s$ and $R_d$. 
Unfortunately, the $B^0_d\to J/\psi \eta'$ branching ratio, which we predict at the 
$5\times10^{-6}$ level, has not yet been measured. But using the other currently available
$B^0_{s,d}\to J/\psi \eta^{(\prime)}$ data in combinations with 
$\mbox{BR}(B^0_d\to J/\psi \pi^0)$, we obtain a picture for the mixing angles 
$\phi_P$ and $\phi_G$ in good agreement with other information. Future measurements of 
$R_s$ and $R_d$ with 30\% precision would result in uncertainties of
$\Delta\phi_P\sim\pm3^\circ$ and $\Delta\phi_G\sim\pm15^\circ$.

We have seen that the amplitude structures of the $B^0_{s,d}\to J/\psi \eta$ and 
$B^0_{s,d}\to J/\psi \eta^{\prime}$ decays correspond formally to the quark--antiquark and
tetraquark descriptions of the $f_0$ in $B^0_{s,d}\to J/\psi f_0$, respectively. 
From the theoretical point of view, the situation in the $B^0_{s,d}\to J/\psi \eta^{(\prime)}$ 
system is more favourable than in $B^0_{s,d}\to J/\psi f_0$ as the hadronic composition of 
the $f_0$ is still not settled. On the other hand, the latter system is more promising from an 
experimental point of view because of the dominant $f_0\to \pi^+\pi^-$ channel.
The most prominent $\eta^{(\prime)}$ decays involve photons or neutral pions in the final 
states, which is a very challenging signature for $B$-decay experiments at hadron colliders
and appears better suited for the future $e^+e^-$ SuperKEKB and SuperB projects, which 
is also reflected by the previous Belle and BaBar analyses of the $B^0_{d,s}\to 
J/\psi \eta^{(\prime)}$ modes. We hope that our experimental colleagues will eventually 
meet the practical challenges, thereby putting yet another system on the roadmap for testing 
the CP-violating sector of the SM, probing non-factorizable $SU(3)_{\rm F}$-breaking effects 
and exploring $\eta$--$\eta'$ mixing.

\newpage

\end{document}